\documentclass[a4paper,superscriptaddress,floatfix,twocolumn]{revtex4}

\usepackage{graphics}
\usepackage{latexsym}
\usepackage{dcolumn}
\usepackage{epsfig}
\begin{document}

\title[The Asakura-Oosawa model in the protein limit]{The Asakura-Oosawa model 
in the protein limit: the role of many-body interactions}

\author{A. Moncho-Jord\'{a} and A.A. Louis} \address{Department of Chemistry, Lensfield Rd,
Cambridge CB2 1EW, UK}

\author{P.G. Bolhuis} \address{Department of Chemical
Engineering, University of Amsterdam, Nieuwe Achtergracht 166, NL-1018
WV Amsterdam, Netherlands}

\author{R. Roth} \address{Max-Planck Institut f{\"u}r 
Metallforschung, Heisenbergstrasse 3, D-70569 Stuttgart, Germany and ITAP,
University of Stuttgart, Pfaffenwaldring 57, D-70569 Stuttgart, Germany}

\begin{abstract}
We study the Asakura-Oosawa model in the "protein limit", where the
penetrable sphere radius $R_{AO}$ is much greater than the hard sphere
radius $R_{c}$.  The phase behaviour and structure calculated with a
full-many body treatment show important qualitative differences when
compared to a description based on pair potentials alone.  The overall
effect of the many-body interactions is repulsive.
\end{abstract}


\maketitle 

\section{Introduction}
The Asakura Oosawa (AO) model\cite{Asak58}, also known as the
penetrable hard sphere  model\cite{Vrij76}, was first introduced
almost 50 years ago to describe depletion effects in colloid-polymer
mixtures.  The colloids are modelled as hard spheres (HS) of radius
$R_c$, and the ``polymers'', or AO particles, as penetrable hard
spheres (PHS), which interact as ideal particles with each other, but
as HS of radius $R_{AO}$ with respect to the colloids.  In spite of
its simplicity, this model has been instrumental in understanding the
phase-behaviour of polymer-colloid mixtures. For example, increasing
the polymer concentration can lead to a fluid-fluid or fluid-solid
demixing of the colloidal particles.  The origin of this phase
transition arises from the depletion effect, which is easily
illustrated by calculating the free-energy of two HS in a bath of
PHS AO particles.  Each HS excludes a volume $\frac43 \pi (R_c + R_{AO})^3$
from the PHS particles, but when two HS approach, their exclusion
volumes overlap, resulting in more free volume available for the PHS
spheres.  This translates into an effective depletion pair potential
between the two particles of the form\cite{Asak58,Vrij76}
\begin{equation}\label{eq1.1}
\beta V_{AO}(r) = -z_{AO}\frac{4 \pi}{3} (\sigma_{cp})^3 \left\{ 1 -
\frac{3}{4} \frac{r}{\sigma_{cp}} +
\frac{1}{16}\left(\frac{r}{\sigma_{cp}}\right)^3 \right\},
\end{equation}in the range $\sigma_{c} < r \leq 2 \sigma_{cp}$,
where $\sigma_{cp} = R_{AO} + R_c$ and $z_{AO}$ is the fugacity of the
AO PHS particles.  The AO potential is always attractive, with a
well-depth that increases with the fugacity (or number density) of AO
particles.  Simple geometrical arguments show that for size ratios
$q=R_{AO}/R_c \leq 0.1547$ there are no higher order many-body
interactions beyond the effective pair potential. For larger $q$,
however, multiple overlap of the depletion zones can occur, leading to
many-body interactions.  As with many soft matter systems, these are
often difficult to calculate, and so a common approximation is to
ignore them, and treat a system in the effective pair-potential
approximation\cite{Loui01a,Liko01}.  For the AO model, this works
surprisingly well for descriptions of phase behaviour up to
size-ratios of $q\approx 1$\cite{Meij94,Dijk99}.

In this paper we study the AO model for $q=R_{AO}/R_c >>1$.  For
polymer-colloid systems this is often called the nano-particle or
protein limit, because small particles such as proteins are needed to
achieve the large size-ratios.  Clearly, a pair potential picture
should break down for large enough $q$, where many-body effects are
expected to dominate. For ideal polymers in the limit $q<1$, the
effect of many-body interaction in polymer and colloid mixture has
been studied by Meijer and Frenkel\cite{Meij94}, who found that these
interactions stabilise the liquid phase. Recent work has shown that
for $q >>1$, many-body effects can qualitatively affect the phase
behaviour\cite{Sear01,Bolh03,foot1}.

 The AO model was originally developed for size-ratios $q < 1$, where
$R_{AO}$ taken to be the radius of gyration $R_g$ of ideal
polymers\cite{Asak58,Vrij76}.  For $q >> 1$ this simple mapping no
longer holds, although the AO model can still be mapped onto a model
for ideal polymers by correctly defining an effective $R_{AO}$
radius\cite{Bolh03}.  However, our goal here is not so much to study
colloid-polymer mixtures, but rather to investigate the effect of
many-body interactions on a well-defined system.  The AO model has
the  particular
advantage  that the effective pair interaction
$V_{AO}(r)$ is exactly known.  By directly calculating the phase
behaviour and structure of a two-component AO model, and comparing it
to an effective one-component model with the AO pair potential, we can
systematically study the effect of many-body interactions.  The
insight gained from this well characterised system should increase our
appreciation of the complexity of many-body effects in soft matter
systems.

The paper is organised as follows: In section 2 we calculate the
phase-behaviour of the AO model, using Monte Carlo (MC) simulations
and several simple theories.   In section 3 we describe the pair
correlation functions, and also the effective colloid-colloid
structure factors.  Finally, we discuss our results in section 4.

\section{Phase behaviour}

\subsection{Monte Carlo Simulations}
We performed Gibbs ensemble Monte Carlo\cite{frenkelbook} simulations
in the semi-grand ensemble, where colloids are treated canonically and
the PHS in the grand canonical ensemble, for three size ratios
$q=3,5,8$.  The total number of colloidal particles ($N=108$) is hence
fixed, although they can exchange boxes. The chemical potential or
fugacity of the PHS is kept constant by the usual grand canonical MC
insertion and extraction moves\cite{frenkelbook}. Further details of
the method can be found in refs.~\cite{Bolh94,Bolh02}.  This setup is
equivalent to an AO mixture in osmotic equilibrium with a reservoir of
only PHS particles\cite{Lekk92}.

\begin{figure}
\centerline{\epsfig{figure=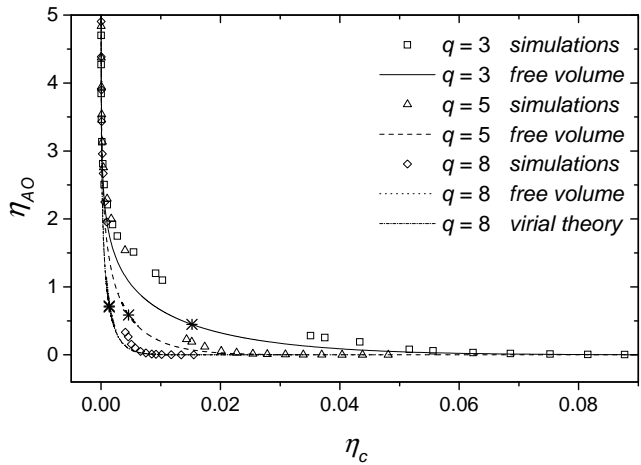,height=6.3cm}}
\caption{\label{PD_LINEAL} Fluid-fluid binodals for size-ratios
$q=3,5,8$. Plotted are the absolute packing fractions $\eta_{AO} =
\frac43 \pi \rho_{AO} R_{AO}^3$ and $\eta_c = \frac43 \pi \rho_c
R_c^3$.   Also shown are the binodals calculated with the free volume
theory of ref.\protect\cite{Lekk92} and their corresponding critical points (asterisks). Note that the critical colloid packing fraction tends to zero with increasing $q$. The free volume and virial theory binodals for $q=8$ can not be distinguished since both theories converge in the limit of large $q$.}
\end{figure}

\begin{figure}
\centerline{\epsfig{figure=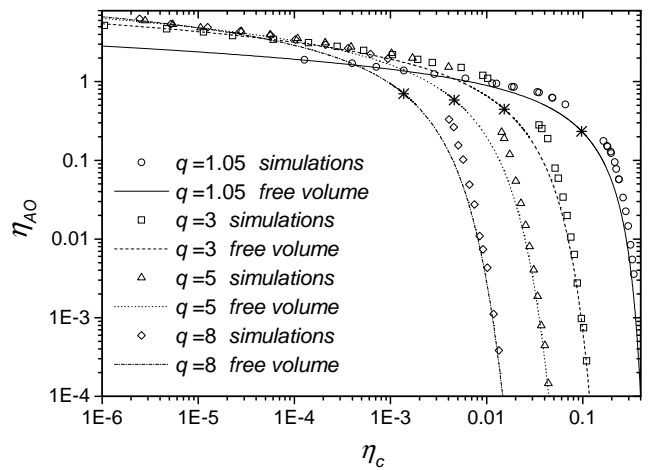,height=6.3cm}}
\caption{\label{PD_LOG} Same as in Fig.~\protect\ref{PD_LINEAL}, but on
a log-log scale.  This emphasizes a different part of the binodals,
showing that the relative agreement of the free-volume theory with the
simulations is about the same for all size-ratios.  Again, the critical points of the free volume binodals are shown as asterisks. Here we also include results for $q=1.05$, taken from ref.~\protect\cite{Bolh02}.}
\end{figure}

Results for the binodals are shown in Fig.~\ref{PD_LINEAL}.  For
increasing $q$, the critical colloid packing fraction $\eta_c =
\frac43 \pi \rho_c R_c^3$ tends to zero, while the PHS packing fraction
$\eta_{AO} = \frac43 \pi \rho_{AO} R_{AO}^3$ increases.  It is also
instructive to compare the binodals on a log-log plot, shown in
Fig.~\ref{PD_LOG}, which emphasises parts of the binodals further from
the critical point.  For example, we see that the binodals cross at
very low $\eta_c$.

\begin{figure}
\centerline{\epsfig{figure=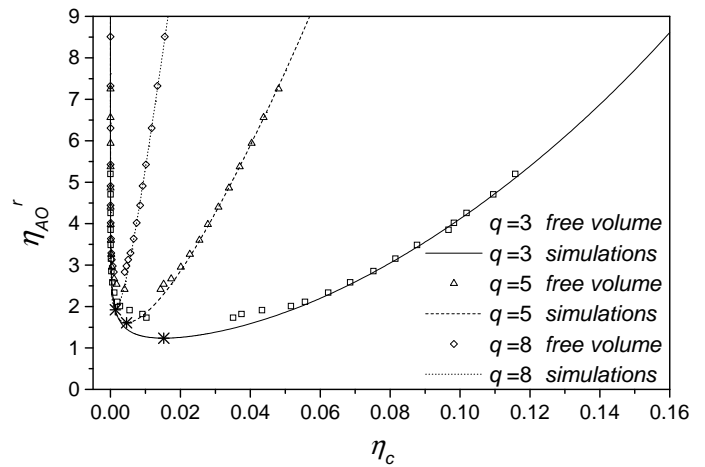,height=6.3cm}}
\caption{\label{PD_RES}
 Fluid-fluid binodals for size-ratios
$q=3,5,8$. The x-axis denotes the colloid packing fraction $\eta_c$ and the 
y-axis the packing fraction $\eta_{AO}^r =
\frac43 \pi \rho_{AO}^r R_{AO}^3$ of a reservoir of pure PHS particles
at the same chemical potential as the two-component AO system.
  Also shown are binodals calculated with the free volume
theory of ref.\protect\cite{Lekk92}, which agree very well with the simulations. }
\end{figure}

It is often convenient to plot the phase diagrams in the semi-grand
ensemble, where the colloids are treated in the canonical ensemble and
the PHS in the grand canonical ensemble.  This is equivalent to
setting up an AO mixture in osmotic equilibrium with a reservoir of
only PHS particles\cite{Lekk92}. Because the PHS particles are ideal,
the fugacity $z_{AO} = \exp(\beta \mu_{AO}) = \rho_{AO}^r$.  The
results in this representation are shown in Fig.~\ref{PD_RES}.

The binodals in Figs.~\ref{PD_LINEAL}~-~\ref{PD_RES} are quite
different to those of the AO model for $q < 1$, shown, for example, in
refs.~\cite{Lekk92,Dijk99}. First of all, they are at much lower
packing fractions of the colloids.  In fact, the critical colloid
packing fraction tends to zero for increasing $q$.  Moreover, in the
reservoir representation, the binodals are much narrower, an effect
that becomes more pronounced with increasing $q$.  Both effects
are similar to what has been found for mixtures of ideal polymers and
colloids\cite{Sear01,Bolh03}.  

The physical origin of the low critical colloid packing fraction is
 the large cross-interaction between the two different
 species. Perhaps a simpler way of viewing this is to first imagine a
 binary HS mixture of particles of equal diameter $\sigma_c$, but with
 a cross diameter $\sigma_{12} > \sigma_c$.  As $\sigma_{12}$
 increases, the system will phase separate at lower and lower volume
 fractions of the HS particles. A phase-diagram like that shown in
 Fig.~\ref{PD_LINEAL} would be symmetric in the packing fractions of
 the two species. However, one could also choose to represent the
 packing fraction of one of the two species, say species $1$, as
 $\eta_1 = \frac{1}{6} \pi \rho_1 \sigma_{12}^3$. For large enough $q$
 the phase diagram would then closely resemble that of
 Fig.~\ref{PD_LINEAL}, since the packing fraction of species $1$ would
 be so low that it could be replaced by an AO PHS particle without
 significantly affecting the phase behaviour.  This way of obtaining
 the large $q$ limit of the AO model also suggests that a simple
 virial theory should become increasingly accurate as $q$ increases
 and the critical packing fractions decrease.  The next section will
 show that such a virial theory works very well indeed\cite{foot2}.

\subsection{Two-component free volume and virial theories}

A general Helmholtz free-energy for the two-component AO model can be
written as:
\begin{equation}\label{eq2.1}
\frac{F(N_c,N_{AO},V)}{V} = f =  f_c^{HS}(\rho_c) + f_{AO}^{id}(\rho_p) + f_{c-AO}(\rho_c,\rho_{AO})
\end{equation}
where the colloids are treated as hard-spheres, and the AO PHS as ideal
particles. We suppress the dependence on temperature $T$, since our
model is athermal.

Whereas $f_c^{HS}(\rho_c)$ and $f_{AO}^{id}(\rho_{AO})$ are well
understood\cite{Hans86}, less is known about the
$f_{c-AO}(\rho_c,\rho_{AO})$ term.  In the free-volume theory of
Lekkerkerker {\em et al.}\cite{Lekk92}, $f_{c-AO}$ is approximated as
\begin{equation}\label{eq2.3}
f_{c-AO}^{free} = \rho_{AO} \omega(\eta_c,q).
\end{equation} In other words, terms proportional to $\rho_{AO}^2$ 
and higher in $f_{c-AO}$ are ignored. $\omega(\eta_c,q)$ can then be
interpreted as the free energy of inserting a single AO PHS particle
into a bath of HS colloids at packing fraction $\eta_c$.  In their
classic paper, Lekkerkerker {\em et al.}\cite{Lekk92} calculated
$\omega(\eta_c,q)$ from scaled particle theory:
\begin{eqnarray}\label{eq2.4}
\omega(\eta_c,q) &=& - \ln(1-\eta_c) + A \frac{\eta_c}{1 + \eta_c}   \\
&+& B
\left(\frac{\eta_c}{1 + \eta_c}\right)^2 + 
C \left(\frac{\eta_c}{1 + \eta_c}\right)^3,  \nonumber
\end{eqnarray}
where $A = 3q + 3 q^2 + q^3$, $B=\frac92 q^2 + 3 q^3$, and $C=3 q^3$.
In the nomenclature of ref.~\cite{Lekk92}, $\omega(\eta_c,q) =
-\ln(\alpha(\eta_c,q))$ with $\alpha(\eta_c,q)$ the so-called
free-volume fraction, which defines the particle packing fraction
$\eta_{AO}^r$ in a reservoir at the same chemical potential as the
mixture i.e. $\eta_{AO} = \alpha(\eta_c,q) \eta_{AO}^r$. 
As $q$ increases, so does the relative strength of the cross
interaction, leading to lower and lower critical colloid packing
fractions.  It  therefore makes sense to
expand $f_{c-AO}^{free}$ in powers of $\eta_c$:
\begin{equation}\label{eq2.5}
f_{c-AO}^{free} = \rho_{AO} \left( \eta_c(1+q)^3 + {\cal O}(\eta_c^2)\right).
\end{equation}
The leading term is proportional to the second cross-virial
coefficient.  In other words, for low $\eta_c$ (and for any
$\rho_{AO}$), free-volume theory reduces, as expected, to a simple
virial theory.

\begin{figure}
\centerline{\epsfig{figure=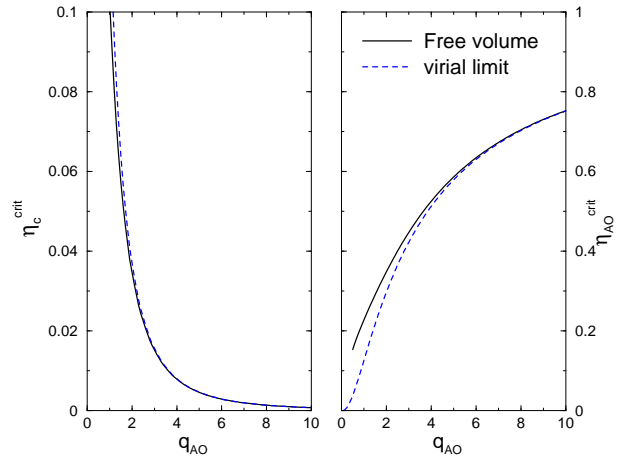,height=6.3cm}}
\caption{\label{etacritAO} The limiting behaviour of the critical point,
given by Eqs.~(\protect\ref{eq2.6a}) and~(\protect\ref{eq2.6b}), is
compared to the full free-volume theory\protect\cite{Lekk92}
calculations.  The differences rapidly decrease with increasing $q$.}
\end{figure}

For the virial theory defined by Eq.~(\ref{eq2.1}) and the leading
term of Eq.~(\ref{eq2.5}), the large $q$ limit of the critical points
can be derived:
\begin{eqnarray}\label{eq2.6a}
\lim_{q \rightarrow \infty} \eta_c^{crit}& = & 
\frac{1}{(1 + q)^3} \sim \frac{1}{q^3} \\ \label{eq2.6b}
\lim_{q \rightarrow \infty} \eta_{AO}^{crit}& = & 
\frac{q^3}{(1 + q)^3} \sim 1 \\\label{eq2.6c}
\lim_{q \rightarrow \infty} \eta_{AO}^{r,crit}& =&  
\frac{\exp(1) q^3 }{(1 + q)^3} \sim \exp(1).
\end{eqnarray}
Since $\eta_c^{crit} \rightarrow 0$ as $q \rightarrow \infty$, it
follows that free-volume theory shows the same limiting behaviour,
which is demonstrated in Fig.~\ref{etacritAO}.

  A comparison with Figs.~\ref{PD_LINEAL}~--~\ref{PD_RES} shows that,
even though higher order $\rho_{AO}$ effects in $f_{c-AO}$ are
ignored, free-volume theory agrees remarkably well with simulations,
just as was found earlier for $q\approx 1$\cite{Dijk02,Bolh02}.  The
main deviations are found near the critical point; these can partially
be ascribed to the fact that free-volume theory is a mean-field
theory, with the wrong critical exponents etc..., which normally leads
to more rounded binodals. Nevertheless, the simulations are quite
consistent with the limiting behaviour for the critical points derived
in Eqs.~(\ref{eq2.6a})~-~(\ref{eq2.6c}).

 In Fig.~\ref{PD_LINEAL}, we also show the binodal for $q=8$ obtained
from the simple cross-virial theory.  The
differences with free-volume theory are not visible on the scale of
the graph.  For smaller $q$ the
virial theory is not quite as good as the free-volume theory, but it
still provides a semi-quantitative description of the binodals,
suggesting that the basic physics can be understood at this simpler
level.

The good agreement between simulations and free-volume theory for the
fluid-fluid binodals suggests that we can use the latter to estimate
the position of the triple point.  This occurs when the gas, liquid
and solid are in equilibrium, and an easy way to estimate the location
of the triple point is to set up an equilibrium between the gas phase
branch of the binodal, and a HS fluid at the freezing transition.  This
results in an approximate position for the triple point at $\eta_c =
0.494$ and:
\begin{equation}\label{eq2.7a}
\eta_{AO}^{r,triple} \approx \frac{\pi}{6}\beta P_{coex} \sigma_c^3 q^3
\approx 6.12 q^3,
\end{equation}
where the reduced coexistence pressure of a HS fluid at freezing,
$\beta P_{coex} \sigma_c^3$, is known from
simulations\cite{frenkelbook}.  In the limit of large $q$ it is
virtually impossible to fit any PHS spheres into the colloidal
crystal, while the gas-phase binodals are at extremely low $\eta_c$;
we therefore expect this relationship to become asymptotically exact
for large size ratios.  In fact, Eq.~(\ref{eq2.7a}) gives a good
prediction for the free-volume triple point  for all
size ratios where a triple point exist, i.e.\ even for $q < 1$.  For
example, at $q=0.8$ free volume theory gives $\eta^r_{AO} = 3.13$
while Eq.~(\ref{eq2.7a}) would predict $\eta^r_{AO} = 3.17$.
This analysis shows that, in the semi-grand ensemble, the triple
point moves to extremely large values of $\eta_{AO}^r$, compared to
the critical point.

\subsection{One-component theory with pair potentials}

It is instructive to compare the results for the two-component AO
model with those of an effective one-component model.  As mentioned in the
introduction, the AO PHS particles can be integrated out to derive an
exact pair potential, valid for all $q$, and given by
Eq.~(\ref{eq1.1}).  For $q \leq 0.1547$ this leads to an exact
description of the system, but for larger $q$, many-body interactions
must be invoked.  Nevertheless, for practical reasons, the pair
approximation is often used in soft matter physics.  In many
situations this works well, but here we expect it to break
down as $q$ increases.

\begin{figure}
\centerline{\epsfig{figure=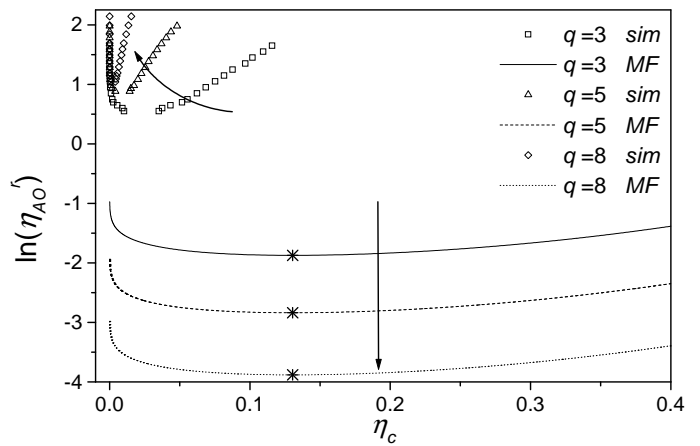,height=6.0cm}}
\caption{\label{SIM_MF} A comparison of the binodals for the
two-component AO model, calculated by GEMC simulations, with an
effective one-component picture, calculated with the pair potential
$V_{AO}(r)$ and Eq.~(\ref{eq2.7}).  Note the qualitative differences
in the limit of large $q$: For the one-component picture
$\eta_c^{crit}$ tends to $0.13$, and $\eta_{AO}^{r,crit}$ tends to
$0$, whereas the full two-component picture shows the opposite
behaviour: $\eta_c^{crit}$ tends to zero, and $\eta_{AO}^{r,crit}$
tends to a constant. These trends are depicted by the arrows. The
differences are due to the effects of many-body interactions.}
\end{figure}

 For large $q$, the pair potential of Eq.~(\ref{eq1.1}) becomes very
long-ranged with respect to the colloidal diameter $\sigma_c$.
Therefore mean-field theory, for which the free-energy takes the form
\begin{equation}\label{eq2.7}
\frac{F^{MF}(N_c,z_{AO},V)}{V}  = f_c^{HS}(\rho_c) + \frac{1}{2} \rho_c^2 \int d{\bf r} V_{AO}(r).
\end{equation}
should become asymptotically exact. For this simple ``van der Waals
limit'', the critical colloid packing fraction is always given by
$\eta_c^{crit} \approx 0.13$, independent of potential details.  Here
we work in the semi-grand ensemble, where the effective pair potential
picture has a consistent statistical mechanical
interpretation\cite{Dijk99,Loui02}.  In Fig.~\ref{SIM_MF}, the
binodals from Eq.~(\ref{eq2.7}) are compared to the full 2-component
simulations.

Besides the obvious quantitative differences for the location of the
critical points (note that the y-axis is logarithmic!), there are
important qualitative differences as well. For example, the
two-component binodals are much more narrow.  Furthermore, the
critical points show opposite scaling behaviour with increasing size
ratio $q$: For the two-component model $\eta_{AO}^{r,crit}$, tends to a
constant and $\eta_c$ tends to zero, whereas for the one-component
model $\eta_{AO}^{r,crit}$ tends to zero and $\eta_c$ tends to a
constant.

The dominant effect of the many-body interactions appears to be
repulsive, since phase-separation occurs at a much higher packing
fraction of the AO PHS particles.  In fact, the three-body
interactions have already been calculated for the AO
model\cite{Goul01}; they are repulsive for all geometries.  This would
seem consistent with the overall effect of the many-body interactions.
However, this interpretation is most likely too naive.  For example,
we expect that the fourth order term is attractive again, and that the
series oscillates, as was recently found in simulations of a self
avoiding walk polymer system\cite{Bolh01a}.  In general, the sign of a
many-body interaction can vary in a complex way with coordinates. A
good example is given by the 3-body HS depletion interactions
calculated in ref.~\cite{Goul01}.  Furthermore, it has recently been
shown that the effects of many-body interactions in a mixture of
interacting polymers and colloids are attractive\cite{Bolh03,Rote03},
the opposite of what we find for the AO model. These examples suggest
that it is generally quite difficult to make simple predictions
regarding the effect of many-body interactions on phase-behaviour.

\section{Pair structure}

\subsection{Radial distribution functions} 

\begin{figure}
\centerline{\epsfig{figure=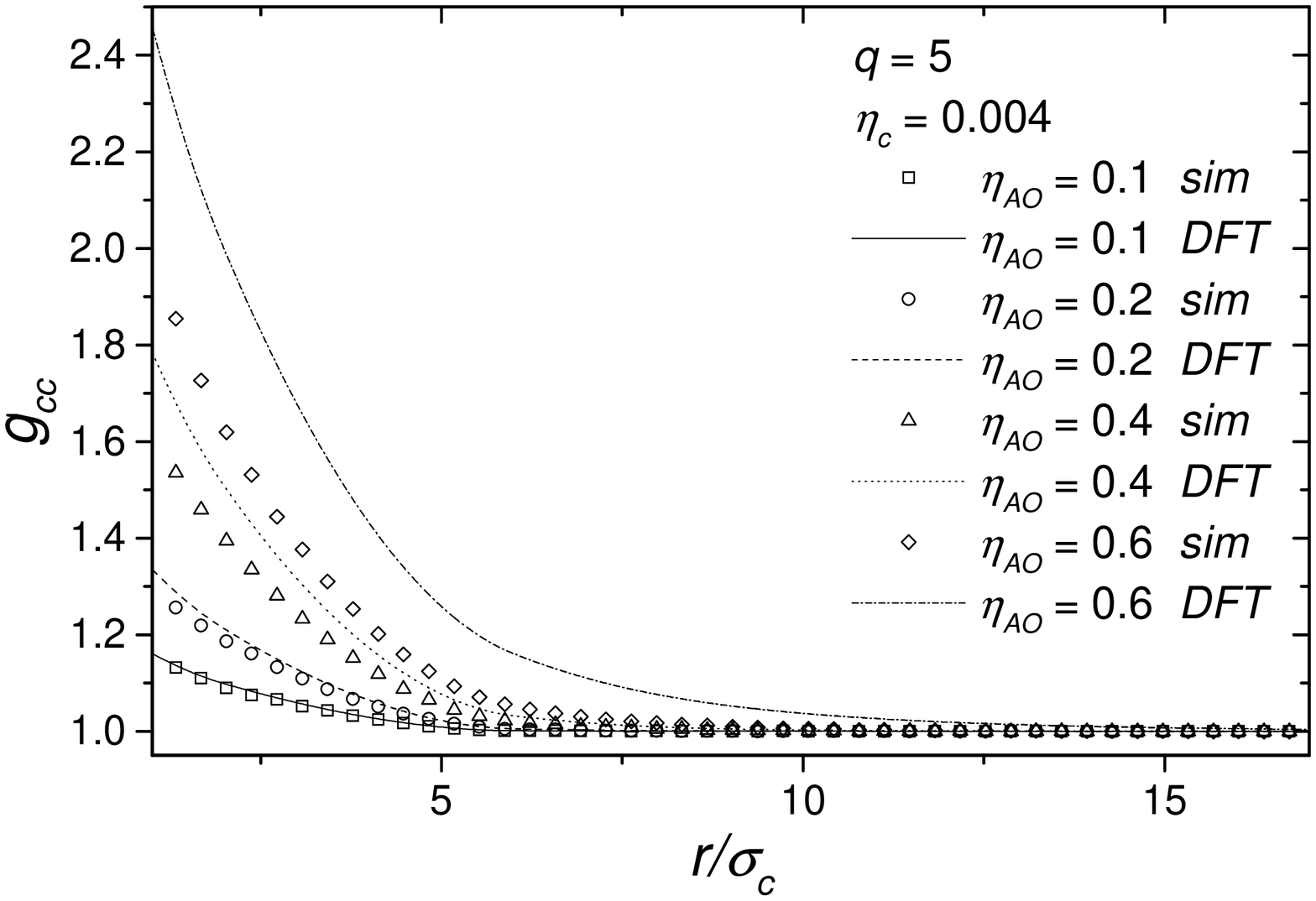,height=6.0cm}}
\centerline{\epsfig{figure=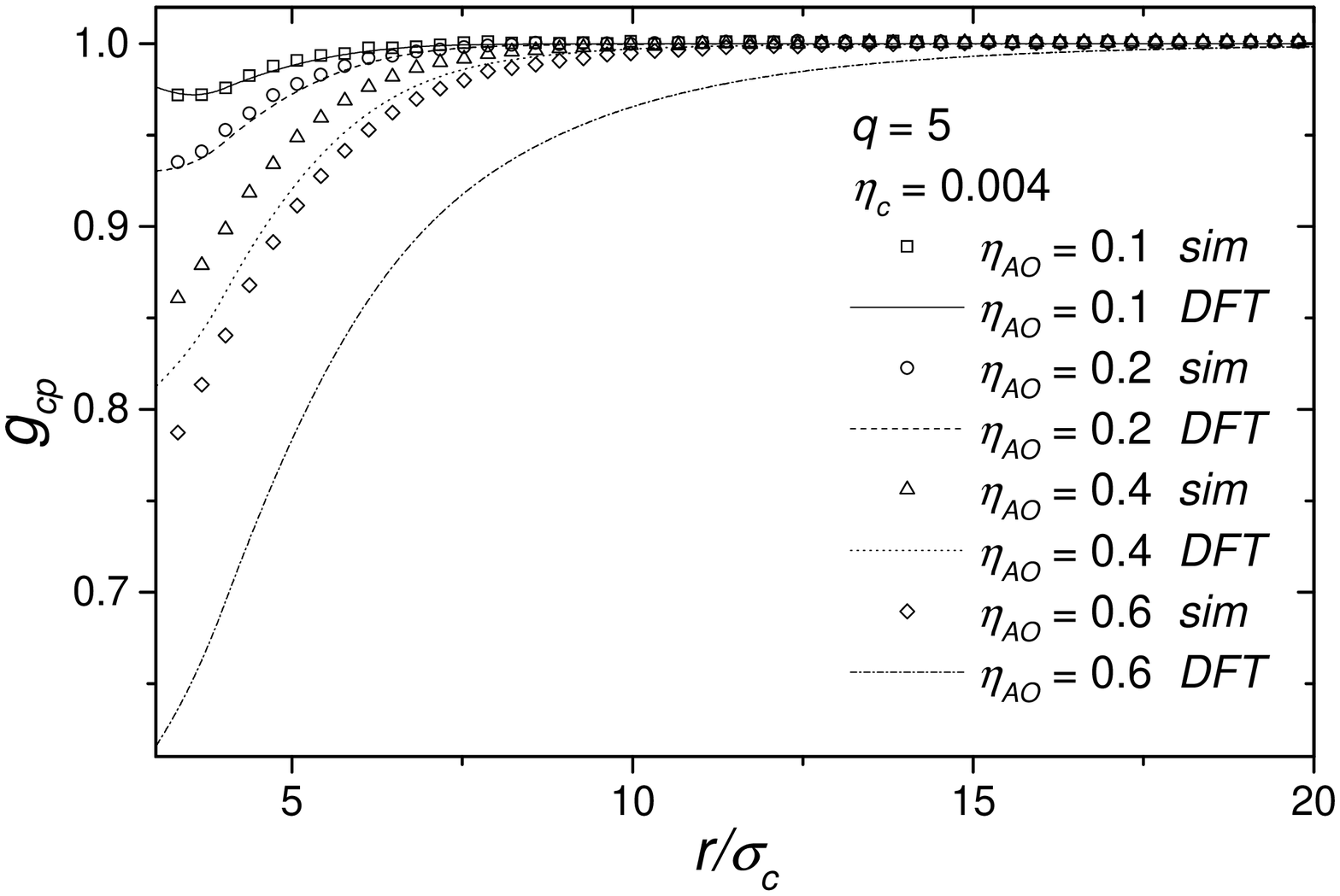,height=6.0cm}}
\caption{\label{GCC_SIM_DFT} The colloid-colloid and colloid-PHS radial
distribution functions for $q=5$ at several state-points.  We compare
direct simulations with results from the test-particle route of a
fundamental measure theory DFT\protect\cite{Schm00}.  For lower
packing fractions the agreement is good, but when the DFT binodal is
approached, differences become larger.}
\end{figure}

Given the unusual phase-behaviour of the AO model in the protein
limit, it should be interesting investigate the effect of many-body
interactions on the pair structure. To that end we performed Monte
Carlo simulations of the colloid-colloid and colloid-AO pair
correlations.  Representative examples are shown in
Fig.~\ref{GCC_SIM_DFT}.  Both $g_{cc}(r)$ and $g_{c-AO}(r)$ have
fairly weak structure, with only one main peak. The reasons for this
are two fold:  1) The colloid packing fraction is very low.  2) More
generally, long ranged interactions lead to less sharply peaked radial
distribution functions\cite{Loui01a}.  We also compared the
simulations to pair correlations derived from the test-particle route
for a recently developed fundamental measure theory (FMT) density
functional theory (DFT)\cite{Schm00}.  This DFT shows the same
phase-behaviour as the free volume theory, which we have shown to be
very accurate for the AO model.  We therefore expect the DFT
to be rather good in this limit, although correlations are often a
more sensitive probe of a DFT than phase behaviour is.  Results are
compared in Fig.~\ref{GCC_SIM_DFT} to the simulations.  For lower
$\eta_{AO}$ the agreement is very good, but for higher $\eta_{AO}$
some deviations are found.  This may partially be because the state
point for the highest $\eta_{AO}$, $(\eta_c = 0.004, \eta_{AO}=0.6)$
is very close to the free-volume critical point (which lies at
$\eta_c = 0.0046, \eta_{AO}=0.586$), while the critical point for the
simulations is further away.

\begin{figure}
\centerline{\epsfig{figure=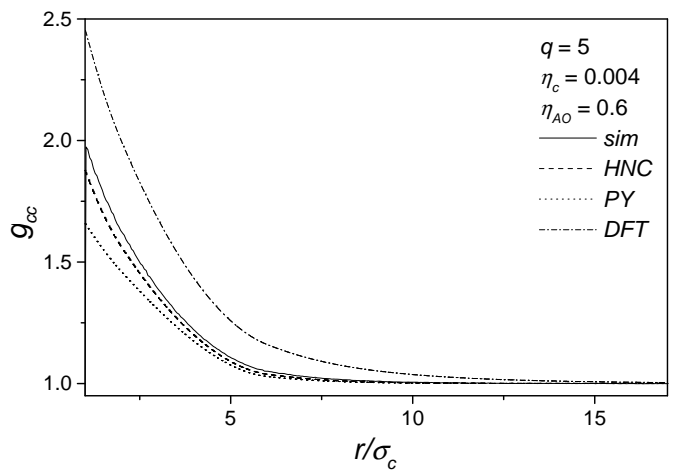,height=6.3cm}}
\caption{\label{GCC_SIM_HNC} A comparison between DFT, HNC, and PY
approaches to the colloid-colloid structure, and results from
simulations.}
\end{figure}

On might expect that, due to the low colloid densities, 
the correlation functions should be rather straightforward to
calculate with integral equation methods\cite{Hans86}.  In
Fig.~\ref{GCC_SIM_HNC} we compare, for one state-point, some
representative results from the Percus Yevick (PY) and Hypernetted
Chain (HNC) approximations\cite{Hans86}.  PY systematically
underestimates the peaks, a general effect that becomes more pronounced
with increasing $\eta_{AO}$, and which is similar to what happens for
binary HS mixtures\cite{Loui01a}.  HNC appears to be more accurate,
but it suffers from a non-solution line, a mathematical artifact where
no solutions are found.  This non-solution line occurs well
before the expected spinodal, and makes HNC less useful.  Finally we
note that direct functional differentiation of the FMT DFT in the
bulk leads to PY correlation functions.  The test-particle route we
apply here is thought to generally be more reliable.

\subsection{Structure factors}

\begin{figure}
\centerline{\epsfig{figure=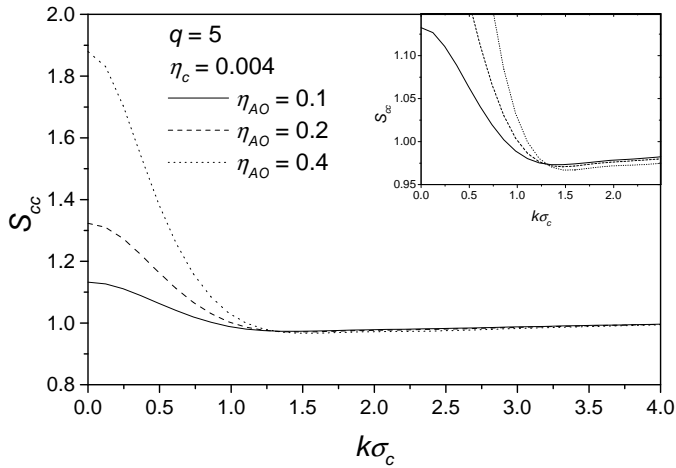,height=6.3cm}}
\caption{\label{SK} Structure factors from simulations for some of  the same
state points as Fig.~\protect\ref{GCC_SIM_DFT}. Inset: the first
isosbestic point where the structure factor is invariant to changes in
$\eta_{AO}$. }
\end{figure}

\begin{figure}
\centerline{\epsfig{figure=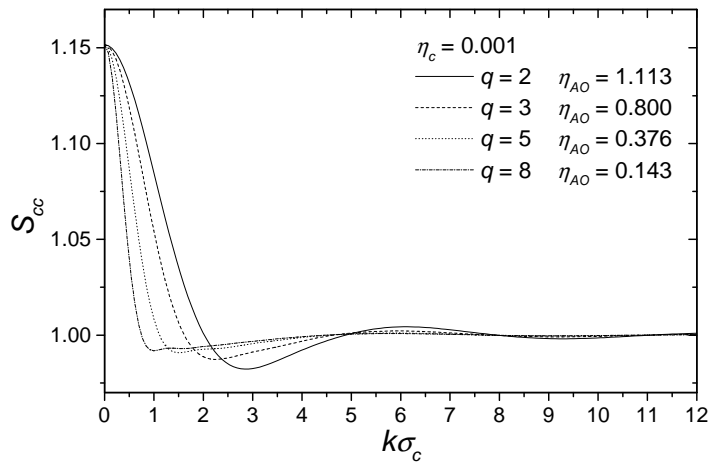,height=6.3cm}}
\caption{\label{SCC_Q} Structure factors from HNC for different values
of $q$.  The state points are chosen to give a similar value of
$S(0)$. The main effect of increasing $q$ is to decrease the value of
$k \sigma_c$ where $S(k)$ begins to increase towards its maximum at
$S(0)$.}
\end{figure}

For scattering experiments, a more useful measure of the pair
correlations is given by the colloid-colloid structure factor
$S(k)$. These are shown in Fig.~\ref{SK} for some of the same state
points as in Fig.~\ref{GCC_SIM_DFT}.  The $S(k)$ shows
virtually no structure, except for a  maximum at $S(0)$.  This
behaviour seems to be generic for the large $q$ limit of the AO model.
Another interesting observation is the occurrence of an isosbestic
point $k_{isos}$, a value of $k$ where $S(k)$ is invariant to changes
in $\eta_{AO}$.  A recent theory\cite{Loui03a} for isosbestic points
predicts that $k_{isos}\sigma_c$ decreases with increasing
range\cite{foot3}. The theory predicts   that $k_{isos} \sigma_c \approx
\pi$ for very short ranged potentials, and that for the AO model 
it can be approximated as:
\begin{equation}\label{eq3.1}
k_{isos} \sigma_c \approx \pi/\left(1 + 0.42 q/2 +  (0.42 q)^2/12\right)
\end{equation}
 Even though this theory was
derived for the small $q$ limit of a one-component model, it still
appears to be semi-quantitative here, since we find an isosbestic
point at $k_{isos} = 1.3 \pm 0.05$ and Eq.~(\ref{eq3.1}) would predict
$k_{isos} = 1.30$.  For other $q$'s we also found good agreement.
This further suggests that one experimental signature of the
long-ranged nature of the AO model is the late upturn of $S(k)$ toward
a maximum at $S(0)$.  This is illustrated in Fig.~\ref{SCC_Q}.  We
note that in a recent paper, Tuinier and Brulet\cite{Tuin03} have
shown similar behaviour from a one-component calculation with only
effective pair potentials. They also performed small angle neutron
scattering experiments on lysozyme-polysaccharide mixtures, measuring
structure factors $S(k)$ that appear quite similar in shape to those
predicted here for the AO model.  Since these authors obtained similar
qualitative behaviour to our simulations from a one-component theory,
the qualitative behaviour of $S(k)$ that we observe is most likely
caused by the long-ranged nature of the pair potentials, and not so
much by the many-body character of the interactions.

\subsection{Effective colloid-colloid pair potentials}

\begin{figure}
\centerline{\epsfig{figure=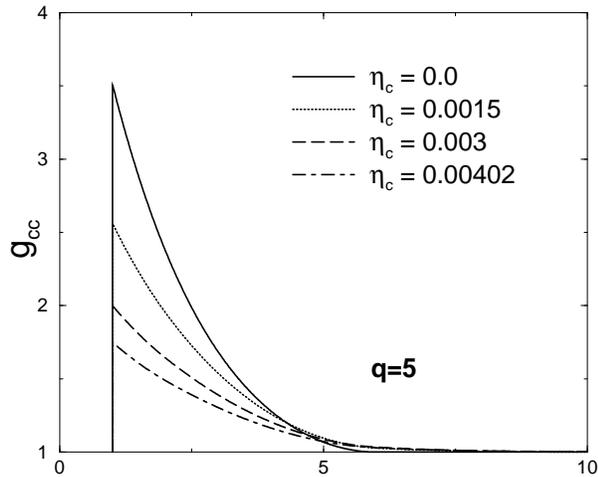,height=6.3cm}}
\caption{\label{gcc_reservoir} Radial distribution functions for
systems with different $\eta_c$.  For each $\eta_c$ the reservoir
packing fraction was $\eta_{AO}^r = 0.964792$, which corresponds to
$\eta_{AO} = 0.4$ at $\eta_c=0.00402$. }
\end{figure}

The extremely low packing fractions of $\eta_c$ studied here might
suggest that one could approximate the pair correlation function by
its zero density limit:
\begin{equation}\label{eq3.3}
\lim_{\eta_c \rightarrow 0} g_{cc}(r) = \exp\left[- \beta V_{AO}(r)\right].
\end{equation}
However, this is not correct, as suggested by the large differences
between the binodals calculated only with $V_{AO}(r)$ and those
calculated for the full two-component AO model.  We demonstrate this
explicitly in Fig.~\ref{gcc_reservoir}, where $g_{cc}(r)$ at the same
$\eta_{AO}^r$ (which means the same $V_{AO}(r)$), but for different
$\eta_c$ was calculated with the FMT DFT approach.  At $\eta_c = 0$,
eq.~(\ref{eq3.3}) is of course exactly obeyed, but this no longer
holds for the other values of  $\eta_c$.  The effect of the many-body
interactions lowers the peak of $g(r)$, which is consistent with the
behaviour of the binodals, where the effect of the many-body
interactions is to reduce the cohesion between the HS colloids.

\begin{figure}
\centerline{\epsfig{figure=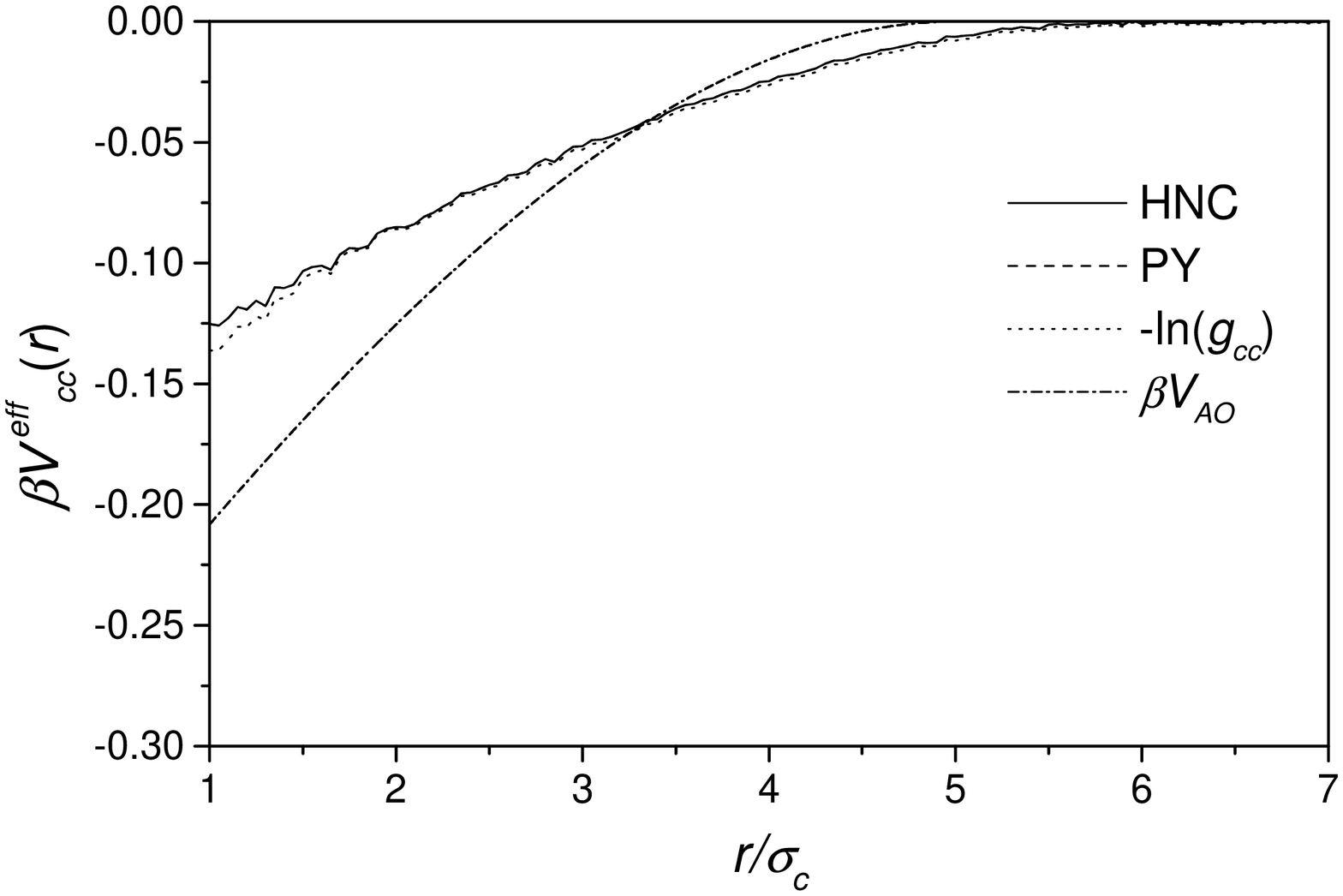,height=6cm}}
\centerline{\epsfig{figure=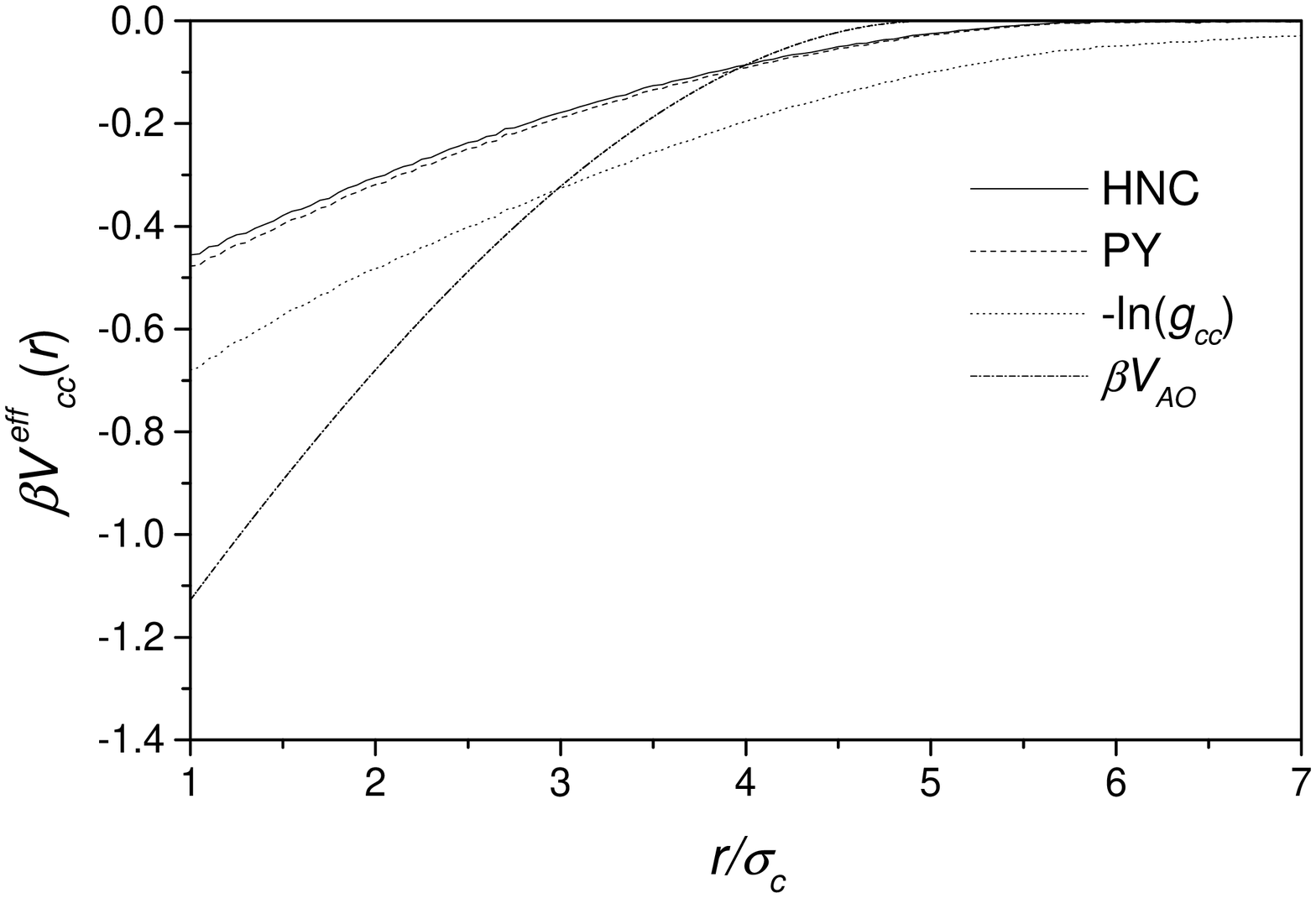,height=6cm}}
\caption{\label{P01_VEFF} Inversions of the colloid-colloid pair
correlation functions for two state-points: up: $\eta_c = 0.0402,
\eta_{AO} = 0.1$ and down: $ \eta_c = 0.0402, \eta_{AO} = 0.6$.  For
each we show the results of a PY and an HNC inversion, as well as the
potential of mean force, $- \ln(g_{cc}(r))$ and the bare two-body 
potential $V_{AO}(r)$.  HNC and PY 
are indistinguishable for the lower polymer packing fraction (upper figure).}
\end{figure}

A general theorem states that for any $g(r)$ and density $\rho$, there
exists a unique pair potential $v_g(r;\rho)$ which will reproduce that
$g(r)$, regardless of the underlying many-body
interactions\cite{Hend74}.  We inverted the $g_{cc}(r)$ at two
state-points with HNC and PY inversions.  The results are shown in
Fig.~\ref{P01_VEFF}. The two inversion methods give very similar
results (especially at low polymer packing fractions, where both inversions 
can not be distinguished on the scale of the graph), suggesting 
that the potential obtained is indeed close to the 
true $v_g(r;\rho_c)$.  Note that $v_g(r;\rho_c)$ is different from the
simple potential of mean force $-\ln(g_{cc}(r))$, which shows that
correlation effects are important here.  As expected, $v_g(r;\rho_c)$
is less attractive than the bare pair potential $V_{AO}(r)$.  It is
also slightly longer ranged.  If there were no many-body forces, then
$v_g(r;\rho_c)$ would be equal to $V_{AO}(r)$ at all densities. The
differences can therefore be attributed to many-body interactions,
whose overall effect is to weaken the effective pair potential.

At each density one could use $v_g(r;\rho)$ to extract the
colloid-colloid pair correlations, and also the osmotic pressure
through the compressibility route.  However, in general a large
density dependence of the effective pair potential also implies
difficulties like the representability issues discussed in
ref.~\cite{Loui02}, making a pairwise description less useful.

\subsection{Effective PHS-PHS pair potentials}

In the limit of large $q$ it might seem tempting to try the opposite
of the usual Asakura-Oosawa strategy, and integrate out the HS
colloids instead of the PHS particles.  This can easily be done, and
the potential between two isolated PHS particles takes the same
functional form as in Eq.~(\ref{eq1.1}), but with colloid and PHS AO
parameters switched.  One difference is that $V_{AO}(r)$ is now
relevant down to $r=0$, because there is no HS repulsion to restrict
it to $r \geq \sigma_c$, as is the case for the colloids.  However, a
description based only on the effective pair potential immediately
leads to problems.  The lack of hard-core repulsion means that the
effective one-component system of PHS particles with an AO type
pairwise attraction falls into the class of catastrophic potentials
defined by Ruelle\cite{Ruel69}, for which there is no thermodynamic
limit.  This does not mean that one cannot derive a consistent
thermodynamics by integrating out the HS colloids.  Rather, because
the PHS particles can overlap so easily, many-body effects are always
very important, especially at small $r$, and are necessary to
stabilise the effective PHS system.  A similar situation was recently
found when integrating out internal degrees of freedom for a solution
of polymers in a poor solvent\cite{Krak03}, as well as for a Gaussian
Core model\cite{Arch02}.  Both examples can lead to catastrophic pair
potentials, even though the underlying many-body system is stable.

\section{Conclusions}

In conclusion then, we have shown that in the so-called protein limit,
where $q >> 1$, the behaviour of the two-component AO model differs
significantly from a description based on an effective pair potential
description alone.   We summarise our main results:
\begin{itemize}

\item In the limit of large $q$, the AO binodals  move to lower and
lower colloid packing fractions $\eta_c$, and at the critical point,
the  AO particle packing
fraction $\eta_{AO}$ tends to a constant.  This is opposite to what is
found from a pair potential description alone, and so this phase behaviour
can be ascribed to many-body interactions.

\item The free-volume theory of Lekkerkerker {\em et al.}\cite{Lekk92}
works remarkably well for the phase-behaviour. It reduces to a simple
virial theory in the large $q$ limit, allowing us to extract limiting
values for the critical points which are consistent with the
simulations.  

\item The colloid-colloid pair-correlation functions show very weak
structure.  Even at very low packing fractions $\eta_c$, $g_{cc}(r)$
is not well described by its zero-density limit $\exp[-\beta
V_{AO}(r)]$.  The pair correlations are well described by a recent
fundamental measure theory density functional theory\cite{Schm00}, at
least if one is not close to the critical point.

\item The overall effect of the many-body interactions is repulsive,
as  seen in the phase behaviour, and also in the structure.

\item There are clear signatures of the long-ranged interactions in
the structure factors $S(k)$.  In particular, the value of $k
\sigma_c$ where $S(k)$ begins to rise to its maximum at $k=0$, decreases
with increasing $q$.

\item A description based on effective pair potentials between the PHS
particles, derived by integrating out the smaller HS colloids, leads
to catastrophic systems with no thermodynamic limit.  
\end{itemize} 

Our aim in this paper was to study the effects of many-body
interactions in a well-defined model system.  Although it would be
tempting to extract some more general insights about the role of
many-body interactions in soft matter systems, this is not so easy to
do. On the one hand, we can make predictions about the behaviour of a
related many-body system, namely a mixture of ideal polymers and HS
colloids in the limit of large $q$\cite{Sear01,Bolh03} where we expect
some similar trends.  But on the other hand, if the ideal polymers are
replaced by interacting ones, the behaviour changes: for example, the
critical colloid packing fraction is almost constant, and the overall
effect of the many-body interactions is attractive instead of
repulsive\cite{Bolh03}.  It is clearly not always easy to predict the
effect of the many-body interactions a-priori.  We conclude from this
that coarse-graining a soft-matter system to a representation where
many-body interactions are important when compared to the pair
interaction may not always be a very fruitful way forward.  Sometimes
it may be easier to treat the original system without this
coarse-graining step.

\section*{Acknowledgements}

A. Moncho-Jord\'{a} thanks the Ram\'{o}n Areces Foundation (Madrid), and 
A.A. Louis thanks the Royal Society (London) for financial support.


\begin{thebibliography}{0}
\bibitem{foot1} These models for ideal polymers
resemble the system studied here, and some preliminary results
were reported in ref. \cite{Bolh03}.
\bibitem{foot2} If one were to keep $R_{AO}$ fixed, and then take 
the limit $q \rightarrow \infty$, then the so-called 
Widom-Rowlinson model\cite{Wido70} results, 
for which a virial theory also works very well\cite{Schm01}
\end{thebibliography}
\end{document}